\documentclass[aps, pre, twocolumn, 10pt,nofootinbib,a4paper]{revtex4-2}
\usepackage{amsmath}
\usepackage{graphicx, tikz, textcomp}
\usepackage{hyperref}
\usepackage[utf8]{inputenc}
\usepackage[T1]{fontenc}
\begin{document}
\title{Charge measurement of $\text{SiO}_2$ nanoparticles in an RF-plasma by IR absorption}
\date{\today}
\author{Harald Krüger}
\email{harald.krueger@physik.uni-greifswald.de}
\author{Elena Thiessen}
\author{Franz Xaver Bronold}
\author{Holger Fehske}
\author{André Melzer}
\affiliation{University of Greifswald, Institute of Physics, Felix-Hausdorff-Str. 6, 17489 Greifswald, Germany}
\begin{abstract}
We have performed measurements of the IR absorption of $\text{SiO}_2$ nanoparticles confined in an argon radio-frequency plasma discharge using an FTIR spectrometer. By varying the gas pressure of the discharge and duty cycle of the applied radio-frequency voltage we observed a shift of the absorption peak of $\text{SiO}_2$. We attributed this shift to charge-dependent absorption features of $\text{SiO}_2$. The charge-dependent shift has been calculated for $\text{SiO}_2$ particles and from comparisons with the experiment the particle charge has been retrieved using our IRPRS (Infrared Phonon Resonance Shift) method. With the two different approaches of changing the gas pressure and altering the duty cycle we are able to deduce a relative change of the particle charge with pressure variations and an absolute estimate of the charge with the duty cycle.
\end{abstract}
\maketitle
\section{Introduction}
\label{sec:intro}
Dusty (or complex) plasmas consist of electrons, ions, neutral gas atoms and additional massive particles. The size of these particles usually ranges between nanometers and $10\,\text{\textmu m}$ in radius. Due to the flux of electrons and ions from the plasma onto the particles, the particles gain a charge. Since the electrons are more mobile, this charge is usually negative in low-pressure, low-temperature laboratory plasmas. The fact that these particles constitute further, observable plasma species has led to the rapid development of the field of dusty plasmas in the last three decades. For overviews, the reader is referred to Refs.~\cite{Verheest2000, Shukla2002, Bellan2006, Piel2017, MelzerBook2019}.
For a long time, the focus of research in this topic has been put on microparticles with the dusty plasma as a model system for atomic or molecular systems due to the rather simple access to properties like the position and velocity of the particles \cite{EThomas2002, Nosenko2006, Feng2007, Williams2011, Buttenschon2011, Himpel2012, Himpel2019a}. However, the dusty plasma community has focussed on the investigation of nanometric particles in a plasma discharge recently. For example, particle growing mechanisms \cite{Hollenstein2000, Kortshagen2016, Boufendi2011}, electron depletion effects \cite{Tadsen2017, Greiner2018} and dust clouds in strong magnetic fields have been investigated \cite{EThomas2012, Tadsen2018}.

Since the particle charge is of fundamental interest for basically every process in dusty plasmas, several approaches have been made to measure it. One example is the resonance method for single particles \cite{Melzer1994, Trottenberg1995, Homann1999, Carstensen2011}. Here, the electric field in the plasma is superposed by an external periodic electric field. From the resonant response of the particles to the changing electric field, the charge can be derived. In another approach thermally excited normal modes are measured in a cluster with a few particles. Comparing the frequency relations of different modes from experiment and theory allows to access the particle charge \cite{Melzer2001, Melzer2003, Qiao2013}. For many-particle systems that support waves the charge can be derived from the analysis of the wave dispersion \cite{Nunomura2002, Nunomura2002PRE, Nosenko2002, Couedel2010}. Nevertheless, all these methods are usually only applicable for dust systems with micron sized particles confined in the plasma sheath where single particle trajectories can be followed by video diagnostics.

Hence, there is a need for a charge diagnostic for (three-dimensional) dust systems with nanoscaled particles. In their approach, \citeauthor{Tadsen2015} exploited dust-acoustic waves in nanodusty plasmas to gain information on the particle charge \cite{Tadsen2015}. However, the presence of waves is not always assured. 

Therefore, \citeauthor{Heinisch2012} have suggested a complementary approach, where the charge-dependent absorption of infrared light by nanoscaled particles is exploited \cite{Heinisch2012, Heinisch2013}. They have shown analytically, that the electrons residing as charges on the particles cause a change in the effective refractive index of the particles. With this change in the refractive index, the light scattering behavior, especially in the infrared wavelength regime of the particles, is shifted and hence, can be used to determine the charge of the particles. \citeauthor{Heinisch2012} have performed calculations for a variety of materials ($\text{Al}_2\text{O}_3, \text{Cu}_2\text{O}, \text{LiF}, \text{MgO}, \text{PbS}$). For these materials it is found that the absorption line of the transverse optical phonon in the infrared spectral range shifts towards higher wavenumbers with increasing particle charge and smaller particle radius. As a rule, the shift is of the order of $10\,\text{cm}^{-1}$ for a particle of 50\,nm radius and a typical charge of about 200\,elementary charges. However, most of the relevant lines for these materials are at wavenumbers far below $1000\,\text{cm}^{-1}$. In principle, the shift should be observable using Fourier transform infrared (FTIR) absorption, but this is a difficult region for FTIR spectrometers. Only aluminum oxide seems to have an absorption in a suitable wavenumber range, however, the required $\alpha$-$\text{Al}_2\text{O}_3$ nanoparticles are difficult to obtain due to technical difficulties in manufacturing processes \cite{Amrute2019}. Another material that is available and has an absorption in a suitable wavenumber range is silicon dioxide ($\text{SiO}_2$). In this work, we demonstrate the charge diagnostic IRPRS (Infrared Phonon Resonance Shift) exploiting the charge-dependent shift of the phonon absorption in experiments and calculations using $\text{SiO}_2$ as a well suited material.

The paper is structured as follows. In the next section we give an overview of the experimental setup. In Sec.~\ref{sec:res} we study the absorption of the nanometric $\text{SiO}_2$ dust in the plasma by FTIR spectroscopy. There, we observe and measure the shift of the phonon absorption with changing plasma conditions. In addition, we perform corresponding calculations of the charge-dependent shift for $\text{SiO}_2$ in Sec.~\ref{sec:calc}. Afterwards, we combine the measurements and calculations to deduce dust charges for the $\text{SiO}_2$ particles. Finally, Sec.~\ref{sec:Conclusion} summarizes our results.

\section{Experimental Setup}
\label{sec:setup}
\begin{figure}
\begin{tikzpicture}[font=\sffamily]
\node[anchor=south west](0,0){\includegraphics[width=\columnwidth]{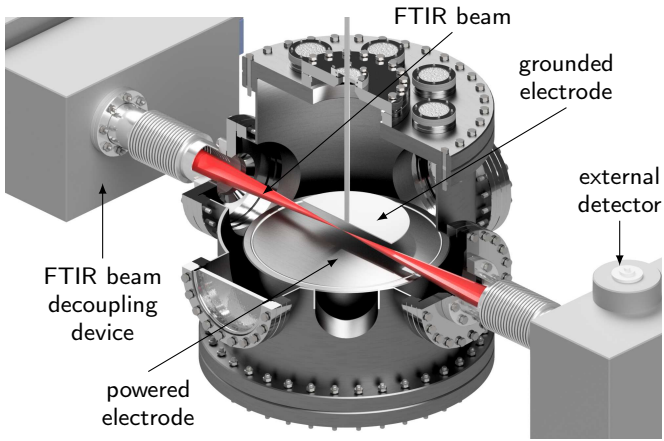}};%
\draw[-latex] (1.4,2.5) node[below, align=center] {FTIR beam\\ decoupling\\ device} -- (1.4,3.5);
\draw[-latex] (8.2,3) node[above, align=center] {external\\ detector} -- (8.2,2.4);
\draw[latex-] (3.5,3.35) -- (6,5.5) node[above] {FTIR beam};
\draw[-latex] (2,1) node[below, align=center] {powered\\ electrode} -- (4.5,2.5);
\draw[-latex] (7.5,4.5) node[above, align=center] {grounded\\ electrode} -- (5,3);
\end{tikzpicture}
\caption{Scheme of the experimental setup with the Bruker Vertex 80v FTIR spectrometer, the FTIR decoupling device, the plasma chamber with one large powered electrode and the FTIR detector. \label{fig:setup}}
\end{figure}
For our IRPRS experiments, we use a low-temperature radio-frequency plasma discharge ignited in a vacuum chamber with an inner diameter of 40\,cm. A lower, powered electrode with a diameter of 30\,cm is separated from the upper, grounded electrode with a diameter of 20\,cm by a distance of 3.5\,cm. Argon is used as a background gas at operating pressures between 4\,Pa and 30\,Pa. The rf generator supplies discharge powers of up to 50\,W. To measure the charge-dependent shift in the infrared a Bruker Vertex 80v FTIR spectrometer is connected to the discharge chamber via two ports, to guide the FTIR beam through the plasma. In addition to the FTIR measurements inside the plasma chamber, the FTIR is equipped with another, internal compartment that is used for measurements of samples that are not being exposed to the plasma environment. The entire FTIR beam paths are under vacuum. A scheme of the experimental setup can be seen in Fig.\,\ref{fig:setup}. For additional information, see \cite{Krueger2021a}.

We run the FTIR spectra at a spectral resolution of $0.08\,\text{cm}^{-1}$. Each FTIR spectrum is averaged over 100 single scans. The measurement time for a spectrum is about 60\,seconds.

We use commercially available silica ($\text{SiO}_2$) nanoparticles of around 40\,nm radius \cite{Goodfellow}. The particles are inserted into the plasma via gas jet injection \cite{Kashu1984, To2009, Krueger2018}. The nanoparticles then form a three dimensional, extended dust cloud where individual particles can no longer be distinguished. The light scattered by the cloud from a sheet of laser light is used to monitor the evolution of the dust cloud.

While other dust materials like PMMA or MF show a significant outgasing and shrinking during long term plasma exposition, $\text{SiO}_2$ particles are stable in the plasma environment and the plasma does not cause a reduction of particle size \cite{Kohlmann2019}.

To reduce the influence of external parameters (optical constants, different response of the internal and external FTIR detectors etc.) and changing clouds we seek for experiments where the charge of the particles is changed using one and the same dust cloud.

As shown by \citeauthor{Ratynskaia2004} \cite{Ratynskaia2004} and \citeauthor{Khrapak2005} \cite{Khrapak2005}, the (absolute value of the) particle charge is seen to be reduced with increasing gas pressure due to stronger ion-neutral collisions that increase the ion flux to the particle. Hence, to change the dust charge within the same cloud in our experiments, we have varied the neutral gas pressure of the plasma. With this, we expect the particle charge to change and to be able to see a change in the IR absorption of the material in the discharge. The IR spectra are measured for neutral gas pressures in the range from 4\,Pa to 30\,Pa.

A second approach is to induce a change in the charge by pulsing the plasma. Thereby, we switch "on" and "off" the plasma at a frequency of 100\,Hz and the particles gain and loose their charge periodically. The frequency of 100\,Hz is low enough to charge and uncharge the particles nearly completely. We have verified using OML (orbital motion limit) model charging current calculations \cite{Cui1994, Nitter1996} that charging and decharging processes are so fast compared to the pulsing frequency that they do not play a role, here. Now, changing the duty cycle of the plasma offers the opportunity to vary the charge. Hence, the temporally averaged charge is nearly proportional to the relative "on"-time-fraction of the plasma. Therefore we can generate a series of different charge numbers for the particles in the plasma and compare the IR spectra with those of the continuously driven plasma.
As expected \cite{Trottenberg1995}, we find only a very minor influence of the plasma power on the absorption.

\section{FTIR measurements}
\label{sec:res}
\subsection{Pressure variations}
\label{sec:res:pressure}
\begin{figure}[!htbp]
	\includegraphics[width=\columnwidth]{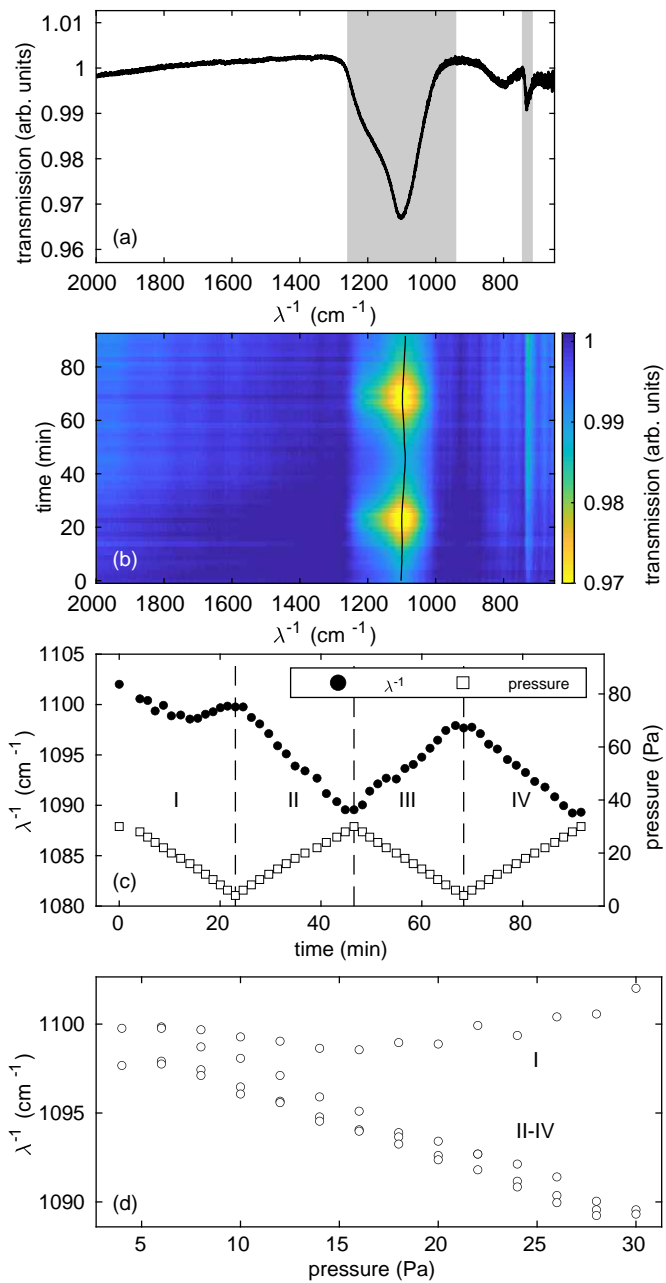}
	\caption{(a) Single transmission spectrum of $\text{SiO}_2$ nanoparticles confined in an argon rf-discharge. (b) Transmission spectra of $\text{SiO}_2$ at a plasma power of 15\,W at different argon gas pressures in the range from 4 to 30\,Pa with position of minimum transmission marked by the black curve. (c) Position of minimal transmission (circles, left axis) and argon gas pressure (squares, right axis) in dependency of the time of the measurement. (d) Position of minimal transmission in dependency of the argon gas pressure.\label{fig:res:press}}
\end{figure}

\begin{figure}
	\includegraphics[width=\columnwidth]{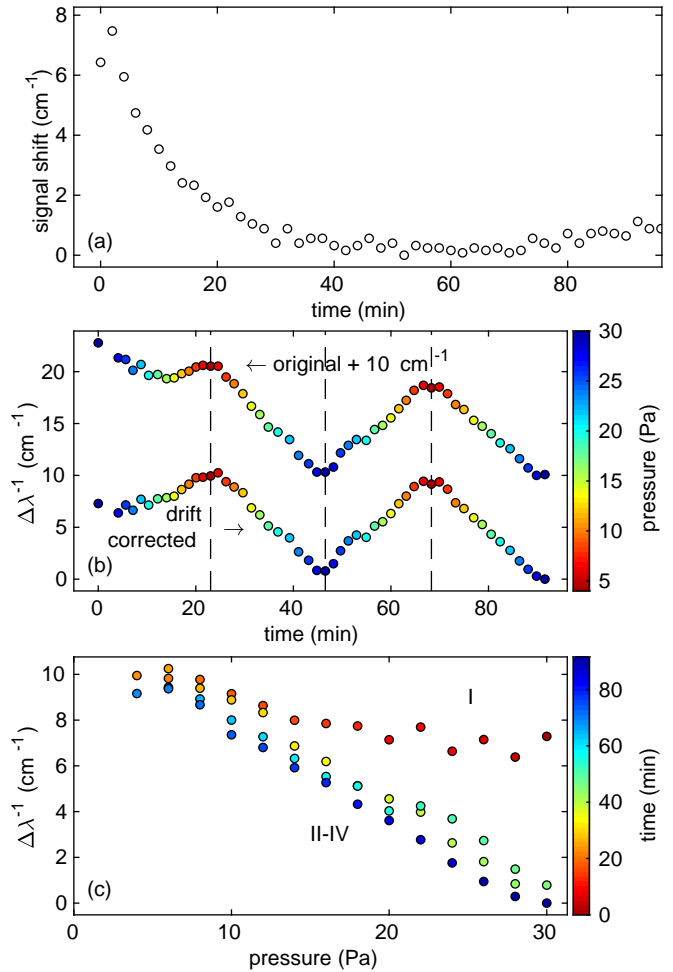}
	\caption{(a) Time resolved shift of the absorption peak position at a constant pressure of 30\,Pa after particle injection. (b) Time resolved change of the particle absorption peak at different pressures (color coded) in original and drift-corrected version. The original time series has been offset by $\Delta\lambda^{-1}=+10\,\text{cm}^{-1}$ for clarity. (c) Corrected shift of the absorption peak position in dependence of the gas pressure.
	\label{fig:res:presstime}}
\end{figure}
We now present the measurements of the FTIR spectra of the confined dust in the plasma environment.
In our experimental approach, we sweep the argon gas pressure from $30\,\text{Pa}$ to $4\,\text{Pa}$ back and forth several times. The measured FTIR absorption spectra are shown in Fig.~\ref{fig:res:press}b. The spectra are displayed in the range from $2000\,\text{cm}^{-1}$ to $600\,\text{cm}^{-1}$. A total of 53 measurements on the same cloud have been done. The absorption amplitude is of the order of a few percent. The interesting absorption structure is located at around $1100\,\text{cm}^{-1}$, where the transmission shows a rather broad local minimum, see Fig. \ref{fig:res:press}a.
In what follows, we focus on this structure (marked by the gray background).

On the one hand, the absolute absorption changes with the changing plasma properties over the measurement time. This can mainly be accounted for by the number of particles in the line of sight of the FTIR beam while the gas pressure changes. On the other hand, we are primarily interested in the wavenumber position of the absorption structure. 

The shape of the absorption peak is found to be constant over time and not depending on the gas pressure. For a reliable measurement, the position of minimal transmission is determined after fitting a polynomial to the absorption signal in the selected spectral range and then using the minimum of the fit (for details on the fit, see Ref.~\cite{Krueger2021a}).

The position of maximum absorption of the charged dust in the plasma depending on the measurement time is displayed in Fig.~\ref{fig:res:press}c together with the corresponding pressure values. In the first pressure scan I (30\,Pa down to 4\,Pa) the absorption minimum shifts from $1103\,\text{cm}^{-1}$ to $1100\,\text{cm}^{-1}$, the second scan II (on the same cloud) from 4\,Pa to 30\,Pa shows a stronger variation of $\lambda^{-1}$, where the absorption feature moves from $1100\,\text{cm}^{-1}$ to $1090\,\text{cm}^{-1}$. The subsequent scans III (30\,Pa down to 4\,Pa) and IV (4\,Pa to 30\,Pa) follow scan II very closely. Obviously, a time-dependent drift of the FTIR signal overlays the measurement spectrum.
Nevertheless, it can easily be seen, that the position of maximum absorption decreases in phases of increasing gas pressure and vice versa, especially for scans II to IV.
Fig.~\ref{fig:res:press}d shows the dependency of the absorption minimum on the argon gas pressure. Apart from the first pressure ramp, the other pressure ramps clearly indicate an increasing wavenumber of the absorption peak with decreasing pressure.

To evaluate the time-dependent drift of the FTIR signal, we perform additional measurements of this process at constant pressure. The IR absorption is measured at a constant argon gas pressure of 30\,Pa every 120\,s for over 90\,minutes. The position of the absorption peak is determined in analogy to the previous measurement. In Fig.~\ref{fig:res:presstime}a, the absorption minimum is shown. As can be seen, the signal position changes strongly in the first few minutes and decreases from about $8\,\text{cm}^{-1}$ at the beginning to roughly zero within 30\,min. (The temporal shift with respect to the minimum value at $t=50\,\text{min}$ is given.) Afterwards, the position of the signal is only weakly shifted.

Now, the original pressure-dependent shift of Fig.~\ref{fig:res:press}c is corrected for the temporal drift (Fig.~\ref{fig:res:presstime}a). In Fig.~\ref{fig:res:presstime}b the original shift at varying pressure and the drift-corrected shift are presented. As can clearly be seen, the characteristic ramping structure with increasing and decreasing pressure can be seen in both the original and the drift-corrected version. Now, the drift-corrected version exhibits increasing shifts $\Delta\lambda^{-1}$ with decreasing pressure also for the first pressure ramp. Nevertheless, the values in the first 20\,min still do not match the values of the next ramping period as nicely as for the following ramping sections. In addition, we have to mention, that due to the drift-correction, we have lost information about the absolute shift and therefore, can only discuss relative changes in the particles IR absorption. The uncharged dust sample measured in the internal sample chamber of the FTIR as mentioned in Sec.~\ref{sec:setup} shows a stable absorption position at $\lambda_0^{-1}=1092.8\,\text{cm}^{-1}$. Nevertheless, due to the drift of the external detectors' signal we cannot uniquely put both measurements in relation to each other. Hence, we decided to show the shift $\Delta\lambda^{-1}$ with respect to the minimum value at 30\,Pa. In Fig.~\ref{fig:res:presstime}c, the drift-corrected shift is shown again in dependency of the applied argon gas pressure. In comparison to Fig.~\ref{fig:res:press}c, also the first 13 measurements in the first ramping section (30\,Pa to 4\,Pa) join in the general trend of an increasing shift with decreasing argon gas pressure.

Before putting this shift of the absorption into relation with a possible change in the charge, we first present another method of changing the particle charge in the plasma.
\subsection{Pulsing}
\label{sec:res:pulsing}
\begin{figure}
	\includegraphics[width=\columnwidth]{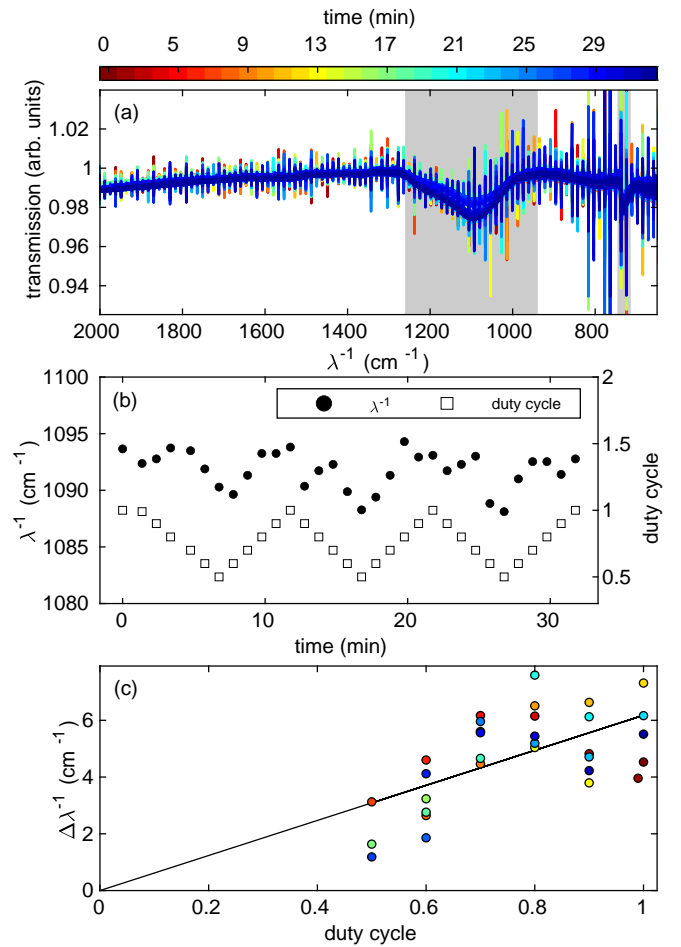}
	\caption{(a) Transmission spectra of $\text{SiO}_2$ nanoparticles confined in an argon rf-discharge at a plasma power of 50\,W at an argon gas pressure of ~10\,Pa at different pulsing duty cycles with a frequency of 100\,Hz. (b) Position of minimal transmission (circles, left axis) and duty cycle (squares, right axis) in dependency of the time of measurement. (c) Shift of the position of minimal transmission in dependence of the duty cycle with linear fit. \label{fig:res:pulse}}
\end{figure}
In this experiment, a new dust cloud is confined in the plasma and we sweep the duty cycle between 50\% and 100\% back and forth. Similar to the pressure variations in Sec.~\ref{sec:res:pressure}, we first show the general spectral overview in Fig.~\ref{fig:res:pulse}a. The spectrum for an applied plasma power of 50\,W at an argon gas pressure of 10\,Pa and duty cycles between 0.5 and 1 looks very similar to the spectrum from the pressure variation. However, in the detail we do find differences: The absolute absorption is a little weaker and the entire spectrum is overlaid with equidistant spikes. Moreover, the spikes depend on the pulsing frequency. Nevertheless, since the spikes have a rather small width, they are clearly separated in the main signal and the absorption minimum is still easily and reliably measurable.

In analogy to the pressure variations, we determine the position of minimal transmission and plot them against the time of the measurement together with the applied duty cycle, see Fig.~\ref{fig:res:pulse}b. The position of minimal transmission ranges between $1095\,\text{cm}^{-1}$ at higher duty cycles and $1090\,\text{cm}^{-1}$ at lower duty cycles and hence, generally follows the ramping of the duty cycle. Although the shift $\Delta\lambda^{-1}$ does not follow exactly the duty cycle, a general trend of decreasing shift with decreasing duty cycle can be determined. Again, we have subtracted the FTIR's time dependent drift of the signal in analogy to Sec.~\ref{sec:res:pressure}.

Basic simulations of the charging and decharging time scales \cite{Cui1994, Nitter1996} show, that the time averaged charge is close to linearly depending on the duty cycle for a frequency of 100\,Hz. We use this expected linear dependence of particle charge with duty cycle to determine an absolute shift $\Delta\lambda^{-1}$ for the data. Therefore, we perform a linear fit to the data and use the extrapolation to duty cycle 0\,\% as the zero point of the shift $\Delta\lambda^{-1}$. From that we get an overall shift of $\Delta\lambda^{-1}=6\,\text{cm}^{-1}$ for a duty cycle of 100\,\%.

As a result, we now have two ways to determine a characteristic shift of the signal, one relative shift with the pressure, and one absolute shift with the duty cycle. Now, to put the shift of the absorption into perspective with the charge, calculations of the resonance shift are presented in the next section.
\section{Theoretical analysis of the resonance shift}
\label{sec:calc}
In this section, we calculate the absorption properties of charged $\text{SiO}_2$ particles following the approach by \citeauthor{Heinisch2012} \cite{Heinisch2012, Heinisch2013}. Due to the positive electron affinity $\chi>0$ of $\text{SiO}_2$, the surplus electrons on the particles cause a polarizability $\alpha = 4\pi\text{i}\sigma_\text{b}/\omega$, where $\sigma_\text{b}$ is the bulk conductivity and $\omega$ is the frequency of the light. This polarizability changes the refractive index of the material to $N=\sqrt{\epsilon + \alpha}$, where $\epsilon$ is the complex dielectric function.
The extinction efficiency $Q_\text{t}$ can be calculated as
\begin{equation}
	Q_\text{t} = -\frac{2}{\rho^2}\sum_{n=1}^\infty(2n+1)\text{Re}(a^r_n+b^r_n)\enspace.\label{eq:Q_t}
\end{equation}
Following the notation of \citeauthor{Bohren1977}, the Mie scattering coefficients can be written as \cite{Bohren1977}
\begin{eqnarray}
a^r_n &= \dfrac{\psi_n(N\rho)\psi_n'(\rho)-N\psi_n'(N\rho)\psi_n(\rho)}{N\psi_n'(N\rho)\xi_n(\rho)-\psi_n(N\rho)\xi_n'(\rho)}\enspace,\label{eq:a}\\
b^r_n &= \dfrac{\psi_n'(N\rho)\psi_n(\rho)-N\psi_n(N\rho)\psi_n'(\rho)}{N\psi_n(N\rho)\xi_n'(\rho)-\psi_n'(N\rho)\xi_n(\rho)}\enspace.\label{eq:b}
\end{eqnarray}
Here, $\rho = 2\pi a/\lambda$ is the size parameter with the radius $a$ of the scattering particle and the wavelength $\lambda$. The Riccati-Bessel functions are defined as
\begin{eqnarray}
	\psi_n(\rho) = \sqrt{\dfrac{\pi\rho}{2}}J_{n+\tfrac{1}{2}}(\rho)\enspace,\\
	\xi_n(\rho) = \sqrt{\dfrac{\pi\rho}{2}}H^{(1)}_{n+\tfrac{1}{2}}(\rho)\enspace,
\end{eqnarray}
with the Bessel function of first kind $J_n(\rho)$ and the Hankel function of first kind $H^{(1)}_n(\rho)$ \cite{Bronshtein2015}. Their derivatives with regard to the argument are marked with a prime.

According to \citeauthor{Spitzer1961} \cite{Spitzer1961}, the complex dielectric function $\epsilon = \epsilon' + \text{i}\epsilon''$ is
\begin{eqnarray}
	\epsilon'(\nu) &= &\epsilon_0 + \sum_j4\pi\varrho_j\nu_j^2\dfrac{\nu_j^2-\nu^2}{(\nu_j^2-\nu^2)^2+\gamma_j^2\nu^2\nu_j^2)}\enspace,\label{eq:eps_r}\\
		\epsilon''(\nu) &= &\sum_j4\pi\varrho_j\nu_j^2\dfrac{\gamma_j\nu\nu_j}{(\nu_j^2-\nu^2)^2+\gamma_j^2\nu^2\nu_j^2)}\enspace,\label{eq:eps_i}
\end{eqnarray}
using an oscillator approach with the strength $\varrho_j$, width $\gamma_j$ and frequency $\nu_j$. In addition, the bulk conductivity becomes
\begin{equation}
	\sigma_\text{b}(\omega) = \dfrac{e^2n_\text{b}}{m^{*}}\dfrac{\text{i}}{\omega+M(\omega)}
\end{equation}
with the elementary charge $e$, the bulk electron density $n_\text{b}$, the conduction band effective mass $m^{*}$ \cite{Heinisch2012}. The memory function is
\begin{equation}
	M(\omega) = \frac{4e^2\sqrt{m^{*}\omega_\text{LO}\delta}(\epsilon_\infty^{-1}-\epsilon_0^{-1})}{3\sqrt{(2\pi\hbar)^3}}\int\displaylimits_{-\infty}^{\infty}\frac{j(-\bar{\nu})-j(\bar{\nu})}{\bar{\nu}(\bar{\nu}-\nu-\text{i}0^+)}\,\text{d}\bar\nu
\end{equation}
with 
\begin{eqnarray}
j(\nu) =&\dfrac{\text{e}^\delta}{\text{e}^\delta-1}|\nu+1|\text{e}^{-\delta(\nu+1)/2}K_1(\delta|\nu+1|/2)\\
\nonumber &+\dfrac{\text{e}^\delta}{\text{e}^\delta-1}|\nu-1|\text{e}^{-\delta(\nu-1)/2}K_1(\delta|\nu-1|/2)\enspace,
\end{eqnarray}
where $\nu = \omega/\omega_{\text{LO}}$, $\delta=\beta\hbar\omega_\text{LO}$, $0^+$ is a small floating point number and $K_1(x)$ is a modified Bessel function (MacDonald function) \cite{Bronshtein2015}.
\begin{table}
	\caption{Optical constants of $\text{SiO}_2$. $\nu_\text{TO}$, $\gamma$, $4\pi\varrho$, $\epsilon_0$ and $\epsilon_\infty$ from \cite{Spitzer1961}, $\lambda^{-1}_\text{LO}$ from \cite{Glinka1997,Lyddane1941} using $m^{*}$ from \cite{Vexler2005} and $\nu_{\text{TO}_7}$.\label{tab:opticalConstants}}
	\begin{ruledtabular}
		\begin{tabular}{clllllll}
		$j$ & \multicolumn{1}{c}{1} & \multicolumn{1}{c}{2} & \multicolumn{1}{c}{3} & \multicolumn{1}{c}{4} & \multicolumn{1}{c}{5} & \multicolumn{1}{c}{6} & \multicolumn{1}{c}{7}\\\hline
		$\nu_\text{TO}$ ($\text{cm}^{-1}$) & 394 & 450 & 697 & 797 & 1072 & 1163 & 1227\\
		$\gamma$ & 0.007 & 0.009 & 0.012 & 0.009 & 0.0071 & 0.006 & 0.11\\
		$4\pi\varrho$ & 0.33 & 0.82 & 0.018 & 0.11 & 0.67 & 0.010 & 0.009\\\hline
		$\lambda_\text{LO}^{-1}$ ($\text{cm}^{-1}$) & 1661.5\\
		$\epsilon_0$ & 2.356\\
		$\epsilon_\infty$ & 4.32\\
		$m^{*}$ & \multicolumn{2}{l}{$0.42\,m_\text{e}$}\\		
		\end{tabular}
	\end{ruledtabular}
\end{table}
\begin{figure}[!htbp]
\includegraphics[width=\columnwidth]{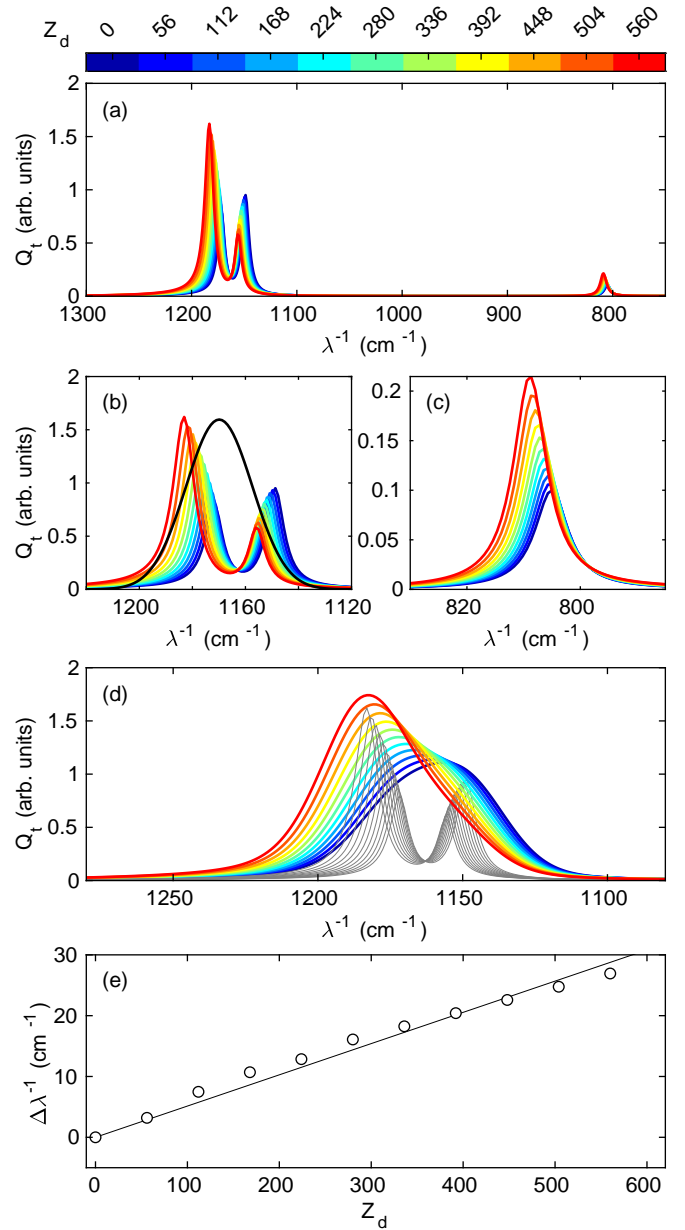}
\caption{Calculation of the extinction efficiency $Q_\text{t}$ of $\text{SiO}_2$ nanoparticles of 40\,nm radius at different charge levels. (a) General overview of the spectra with extinction peaks at around $1180\,\text{cm}^{-1}$, $1150\,\text{cm}^{-1}$ and $800\,\text{cm}^{-1}$. (b) Detailed view of the spectra in the spectral range from $1250\,\text{cm}^{-1}$ to $1100\,\text{cm}^{-1}$ and the used Gaussian (black, not to scale in $Q_\text{t}$-direction). (c) Detailed view of the spectra in the spectral range around $800\,\text{cm}^{-1}$.  (d) Convolution of the spectra and the Gaussian from (b) [colored, not to scale in $Q_\text{t}$-direction. Gray curves show the original spectra as in (b)]. (e) Shift of the maximum position of the convolution $\Delta\lambda^{-1}$ in dependence of the particle charge $Z_\text{d}$.\label{fig:calc:Q_t}}
\end{figure}

Now, we use the constants from Tab.~\ref{tab:opticalConstants} to calculate the dielectric function with \eqref{eq:eps_r} and \eqref{eq:eps_i}. The interesting spectral range is around $1100\,\text{cm}^{-1}$. The bulk conductivity is calculated using the bulk electron density. From OML theory \cite{Mott-Smith1926, MelzerBook2019} as an upper bound, the number of electrons $Z_\text{d}$ on our 40\,nm radius $\text{SiO}_2$ particles can be estimated to be of the order of a few hundred elementary charges. Hence, for the analysis we assume the charges to be in the range of $0 \leq Z_\text{d} \leq 600$.
The bulk electron density then is
\begin{equation}
	n_\text{b} = \frac{Z_\text{d}}{V}=\frac{3Z_\text{d}}{4\pi a^3}\enspace.
\end{equation}
With that, we now determine the polarizability and the refractive index $N$ of $\text{SiO}_2$ using \eqref{eq:a} and \eqref{eq:b} to obtain the scattering coefficients and finally using \eqref{eq:Q_t} to derive the extinction efficiency $Q_\text{t}$.

The calculated extinction efficiencies are shown in Fig.~\ref{fig:calc:Q_t}a for various charge numbers on the particle. A major extinction feature is found in form of two peaks at around $1180\,\text{cm}^{-1}$ and $1150\,\text{cm}^{-1}$. In addition, a rather small extinction is present at around $800\,\text{cm}^{-1}$ (more detailed in Fig.~\ref{fig:calc:Q_t}c).

It can be seen, that all extinction structures change with the charge on the particles towards larger wavenumbers as expected \cite{Heinisch2012,Heinisch2013}. The more detailed view in Fig.~\ref{fig:calc:Q_t}b shows the two peaks around $1180\,\text{cm}^{-1}$ and $1150\,\text{cm}^{-1}$. Both peaks shift with dust charge. Moreover, there also is a change of the amplitude of the absorption efficiency with dust charge. While the amplitude of the extinction peak at around $1150\,\text{cm}^{-1}$ decreases with the charge, the amplitude of the peak at $1180\,\text{cm}^{-1}$ increases with the charge. The shift of the peaks is in the range of about $7\,\text{cm}^{-1}$ for the $1150\,\text{cm}^{-1}$ extinction and about $11\,\text{cm}^{-1}$ for the $1180\,\text{cm}^{-1}$ extinction when changing the charge from 0 to 600 elementary charges.

Comparing the calculated (Fig.~\ref{fig:calc:Q_t}) and measured (Fig.~\ref{fig:res:press}) spectra, we see some differences. In our measurements, we only detect a single, rather broad absorption peak around $1100\,\text{cm}^{-1}$, while in the calculations we see two peaks around $1180\,\text{cm}^{-1}$ and $1150\,\text{cm}^{-1}$. The width of the measured  absorption feature is about $250\,\text{cm}^{-1}$, while the calculated peaks are much narrower with a width of $15\,\text{cm}^{-1}$ to $25\,\text{cm}^{-1}$. 

Possible reasons for the mentioned differences might be the limited resolution of the FTIR spectrometer, a size (and thus charge) distribution of the dust, material inhomogeneities etc. which all lead to a broadening of the signal.

To take these factors into account, it seems reasonable to convolute the calculated peaks with a Gaussian as indicated in Fig.~\ref{fig:calc:Q_t}b. The convolution of the calculated extinction efficiencies $Q_\text{t}$ and the Gaussian can be seen in Fig.~\ref{fig:calc:Q_t}d. The Gaussian convoluted calculations now feature a single peak. The convoluted calculations not only react to the individual shift of the two components, but also the change of the amplitudes of the constituent peaks. The shift of the maximum position of the convolution is now larger than the single-peak shifts due to the different change of the amplitudes of the two peaks.

Hence, $\text{SiO}_2$ provides us with the fortunate situation of two close peaks with a charge-dependent changing amplitude, that in the convoluted amplitude results in a larger shift than the two peaks individually. With this combined effect the charge-dependent shift becomes measurable even for the broad peak seen in the experiment.

We finally determine the maximum position of the calculated, Gaussian convoluted combination of the two absorption peaks with changing dust charges, see Fig.~\ref{fig:calc:Q_t}e. The maximum shifts from $1156\,\text{cm}^{-1}$ at zero charge to $1183\,\text{cm}^{-1}$ at 560 elementary charges. We see a nearly linear relation between charge and shift, and we find a relation of charge and the relative charge-dependent shift of
\begin{equation}
	\Delta\lambda^{-1}(Z_\text{d}) = 0.0514\,\text{cm}^{-1}\cdot Z_\text{d}\label{eq:fit}\enspace.
\end{equation}
From that, now the charges on the $\text{SiO}_2$ particles can be retrieved.

In our measured FTIR spectra, the absorption peak of interest shows up around $1100\,\text{cm}^{-1}$, and another smaller peak at around $730\,\text{cm}^{-1}$, see Fig.~\ref{fig:res:press}a. The calculations show peaks at around $1175\,\text{cm}^{-1}$, see Fig.~\ref{fig:calc:Q_t}d and $804\,\text{cm}^{-1}$ see Fig.~\ref{fig:calc:Q_t}c. We see, that the measured spectra are shifted with respect to the calculations by about $72\,\text{cm}^{-1}$. The calculations of the extinction efficiency are sensitive to the optical constants mentioned in Tab.~\ref{tab:opticalConstants}. The optical constants could be adjusted to match the observed position of the absorption features. However, the absolute position is not of such a vital importance as the relative shift due to the charging.
\section{Extracting the particle charges}
\label{sec:evaluation}
\subsection{Pressure variation}
\label{sec:evaluation:press}
\begin{figure}
	\includegraphics[scale=1]{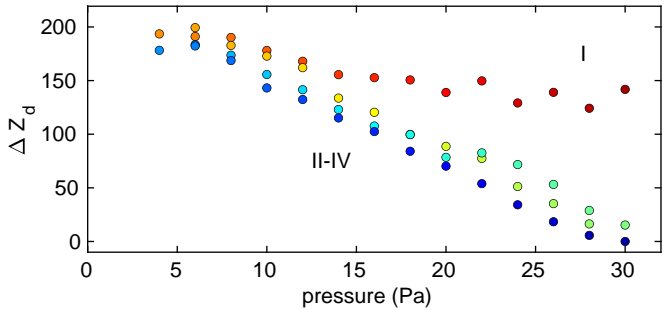}
	\caption{Derived change of the particles charge as a function of the gas pressure. The plasma power is $P=15\,\text{W}$. \label{fig:res:charge}}
\end{figure}
Returning to the measurements, we now will combine the calculated shifts with the experimental results to extract the dust charge. From the shift of the measured IR absorption $\Delta\lambda^{-1}$ in Fig.~\ref{fig:res:presstime}c, we find the relative dust charge $\Delta Z_\mathrm{d}$ from \eqref{eq:fit} as shown in Fig.~\ref{fig:res:charge}. As mentioned above, we have set the position of zero shift $(\Delta\lambda^{-1}=0)$ to the minimum at 30\,Pa. The relative charge difference in comparison to the particle charge at 30\,Pa is about 200 elementary charges at 4\,Pa. For decreasing argon gas pressures, the relative charge increases nearly linearly. The ramping runs II to IV reproduce very similar results.

Measurements from \citeauthor{Ratynskaia2004} \cite{Ratynskaia2004} and calculations from \citeauthor{Khrapak2005} \cite{Khrapak2005} showed a clear inverse proportional dependence of the particles charge on the neutral gas pressure ($Z_\text{d}\sim 1/p$). We are only investigating relative charge differences with respect to our measurement at 30\,Pa, while the other studies determined absolute charges. Hence, in our measurements such a clear $1/p$ dependency cannot be revealed. In addition, \citeauthor{Ratynskaia2004} investigated on a pressure range from 20\,Pa to 100\,Pa and \citeauthor{Khrapak2005} between 20\,Pa and 150\,Pa, while our measurements take account for the pressure range from 4\,Pa to 30\,Pa. Therefore, our used pressure range is only somewhat overlapping with the other studies.

With these differences between the mentioned studies and our experiments it is hard to compare the results in detail. Nevertheless, the general trend of decreasing particle charges with increasing gas pressure can be confirmed.

\subsection{Pulsed plasma}
\label{sec:evaluation:pulse}
\begin{figure}
	\includegraphics[scale=1]{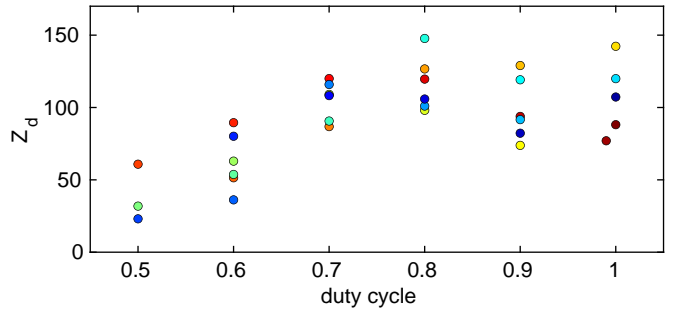}
	\caption{Derived charge of the particles confined in the pulsed plasma environment at different duty cycles. The argon gas pressure is $p=10\,\text{Pa}$ and the plasma power $P=50\,\text{W}$.\label{fig:res:charge_dc}}
\end{figure}

Since we expect a linear dependence of duty cycle and time averaged charge, the charge of the particles can be determined absolutely by assuming that at duty cycle 0\,\% the charge on the particles is $Z_\text{d}=0$, see Figs.~\ref{fig:res:pulse} and~\ref{fig:res:charge_dc}. The resulting charge ranges for the particles in the continuously driven plasma environment (duty cycle of 1) between 100 and 150 elementary charges. With reduced duty cycle, the determined charge goes down to about 50 elementary charges at a minimal applied duty cycle of 0.5. Concerning the accuracy of the results it has to be mentioned, that the values scatter rather strongly with ranges of up to 80 elementary charges.

\subsection{Discussion}
To further put our IRPRS results into perspective, we now study the particle charge from OML theory \cite{Mott-Smith1926, MelzerBook2019}. We assess the electron temperature to be of the order of $T_\text{e}=3\,\text{eV}$ \cite{Klindworth2006}. This results in a charge of 168 elementary charges for our particles of 40\,nm radius. With respect to our charge measurements, this fits rather fine into the range of the charges of the continuously driven plasma in the pulsing approach, see Fig.~\ref{fig:res:charge_dc}.

Considering ion-neutral collisions in the charging process, a change of the particle charge of about 40 elementary charges in the pressure range from 4\,Pa to 30\,Pa would be expected following \citeauthor{Khrapak2005} \cite{Khrapak2005}. This is about half a magnitude smaller than our derived change in the particle charge.

Nevertheless, in such dense plasmas electron depletion also known as the Havnes effect \cite{Havnes1987} becomes important. This effect especially occurs in dusty plasmas with submicron to nanometer sized particles, where a large fraction of the free electrons is bound to the particles \cite{Tadsen2015,Tadsen2017,Krueger2018}. From the absolute transmission of the FTIR signal, we have deduced a line-averaged dust density of maximum $n_\text{d}=2\cdot10^{13}\,\text{m}^{-3}$. Compared to other experiments with larger nanoparticles, this is a rather small density \cite{Tadsen2015,Krueger2018}. Using this dust density and the measured dust charge, the Havnes parameter \cite{Havnes1987, MelzerBook2019} can be estimated to be $P\approx3$, indicating only little electron depletion in our experiments in contrast to experiments in dense dust clouds with $10\leq P\leq50$ \cite{Tadsen2015}.

We now want to compare both of our measurements. From the experiment with varying  duty cycles of the plasma power we have derived a charge of about 150 elementary charges at an argon gas pressure of 10\,Pa. From our pressure variation measurement, a relative charge difference of about 150 elementary charges at 10\,Pa in comparison to 30\,Pa can be estimated. With regard to the uncertainties that can clearly be seen in the duty cycle variation measurements (Fig.~\ref{fig:res:charge_dc}), both results are generally in agreement with each other. Nevertheless, combining both results and extrapolating an absolute charge at 30\,Pa by taking $Z\approx170$ at 10\,Pa and subtracting the relative charge change between 10\,Pa and 30\,Pa of $\Delta Z \approx 150$ would imply nearly no charges at argon gas pressures of 30\,Pa. In Ref.~\cite{Khrapak2005} a dependency of the charge number on the pressure  
of $Z_\text{d}\sim 1/p$ is suggested. Our pressure-dependent dust charges would follow such a dependency when the absolute charge at 30\,Pa would be $Z_\text{d}\approx50$.

Especially with the rather large scatter of the duty cycle variations both approaches (variation of the pressure and variation of the duty cycle) still seem to be reasonably compatible with each other.

Furthermore, in the pressure variation approach, the applied plasma power is only 15\,W, while for the pulsing a power of 50\,W at continuously driven plasma has been applied. The high plasma power is necessary in the pulsing approach to have a stable confinement especially at small duty cycles. In contrast, the lower plasma power at the pressure variation is necessary for a stable confinement over the wide gas pressure range from 4\,Pa to 30\,Pa. Although the plasma power is not expected to change the electron temperature that drastically, the measurement conditions are still different. Therefore, the different conditions are additional potential factors for differences in the results of both methods.

In addition, it has been reported, that the particle bulk temperature can also play a role on the absorption properties. In the study \cite{Krueger2021a} the temperature of melamine-formaldehyde particles has been investigated in dependence of plasma exposure by FTIR spectrometer measurements. It has been shown, that with increasing temperature, some of the particles absorption peaks shift to lower wavenumbers, thus in the opposite direction of the shift of increasing charges. This might lower the shift caused by the charge of the particles. We have tested the temperature dependence of the absorption and did not find a significant dependence in a reasonable temperature range for the $\text{SiO}_2$ particles.

However, comparing our IRPRS method with other charge diagnostics we still want to mention some advantages. As mentioned in Sec.~\ref{sec:intro}, a hand full of techniques for charge measurements are available and have been widely used for microparticles. But when it comes to nanodusty plasmas the diagnostics rely on the existence of dust-density waves in the plasma. With our approach not only systems without dust-density waves are examinable, but the diagnostic is also independent of the knowledge of other plasma parameters like the electron temperature or the particle densities of the several plasma species. The IRPRS method generally is an absolute method to determine the particle charge in a nanodusty plasma, although it yields line-averaged data.
Nevertheless, the technique is not applicable to all types of material, yet. The advantage of rather well known optical properties of silica allows us to develop this diagnostic for silica nanoparticles. In contrast, typical particle materials (e.g. melamine formaldehyde) are lacking the availability of the necessary parameters.

Besides laboratory plasmas the approach might also be useful for diagnostics in astrophysical dusty plasmas. The optical, non-invasive access to the particle charge by investigating the shift of the absorbing phonon resonance is a great advantage. In addition, the non-invasivity of our method can be of high interest especially for industrial manufacturing processes.

\section{Conclusions}
\label{sec:Conclusion}
To sum up, we have measured the IR absorption of $\text{SiO}_2$ nanoparticles of 40\,nm radius confined in a plasma. Plasma conditions (pressure and duty cycle of the RF power) have been varied to cause a change of the particle charge. We have observed a clear change in the position of the particles' major absorption peaks with changing plasma conditions. Calculating the extinction efficiency of silica nanoparticles with respect to changes in the charge of the particles, we found a shift of the major absorption peaks towards higher wavenumbers with increasing particle charge. Combining experiment and theory, we were able to deduce a relative change in dust charge with changing gas pressure. By varying the duty cycle it was even possible to estimate the absolute charge of the particles. In general, the results of this novel approach of a charge estimation are in good agreement with OML theory and other experimental charge measurements. In conclusion, this technique provides us with a non-invasive charge diagnostic that is applicable to nanodusty plasmas.

\begin{acknowledgments}
We would like to thank P. Druckrey for the technical support in the preparation phase of the experiments. We gratefully acknowledge previous work from R.L. Heinisch which provides a basis for the current study. This work was financially supported by the Deutsche Forschungsgemeinschaft via Project DFG 1534 Me8-1. 
\end{acknowledgments}

\end{document}